\documentclass[sigconf]{acmart}

\PassOptionsToPackage{obeyspaces}{url}
\AtBeginDocument{%
  \providecommand\BibTeX{{%
    \normalfont B\kern-0.5em{\scshape i\kern-0.25em b}\kern-0.8em\TeX}}}

\copyrightyear{}
\acmYear{}
\acmConference[]{}{}{}
\acmBooktitle{}
\acmDOI{}
\acmISBN{}
\renewcommand\footnotetextcopyrightpermission[1]{} % removes footnote with conference information in first column

\usepackage{tablefootnote}
\begin{document}

%%
%% The "title" command has an optional parameter,
%% allowing the author to define a "short title" to be used in page headers.
\title{Performance Bottlenecks in Computing ZK Proofs on a GPU}

%%
%% The "author" command and its associated commands are used to define
%% the authors and their affiliations.
%% Of note is the shared affiliation of the first two authors, and the
%% "authornote" and "authornotemark" commands
%% used to denote shared contribution to the research.

\author{Victor Carolino}
\authornote {Work done while these authors were at Lehigh University}
\email{victor.carolino@proton.me}
\affiliation{
\institution{Lehigh University}
\city{Redwood Shores}
\state{CA}
\country{USA}
}
\author{Caleb Geren}
\authornotemark[1]
\email{c.geren@utah.edu}
\affiliation{
\institution{University of Utah}
\city{Salt Lake City}
\state{UT}
\country{USA}
}
\author{Michael Kaufman}
\authornotemark[1]
\email{kaufmanmp03@gmail.com}
\affiliation{
\institution{Lehigh University}
\city{Burlington}
\state{MA}
\country{USA}
}
\author{Maximilian Nussbaum}
  \email{nussbaummax88@gmail.com}
\affiliation{%
  \institution{Lehigh University}
  \city{Bethlehem}
\state{PA}
\country{USA}
  }
\author{William Scott}
\email{wis226@lehigh.edu}
\affiliation{%
  \institution{Lehigh University}
    \city{Bethlehem}
\state{PA}
\country{USA}
}
\author{Henry F. Korth}
\email{hfk2@lehigh.edu}
\affiliation{%
  \institution{Lehigh University}
    \city{Bethlehem}
\state{PA}
\country{USA}
}

%%
%% By default, the full list of authors will be used in the page
%% headers. Often, this list is too long, and will overlap
%% other information printed in the page headers. This command allows
%% the author to define a more concise list
%% of authors' names for this purpose.
\renewcommand{\shortauthors}{Carolino, et al.}

%%
%% The abstract is a short summary of the work to be presented in the
%% article.
\begin{abstract}
Zero-knowledge proofs allow one party to prove the correctness of a statement or execution of a computation while revealing no more than the statement of correctness itself. Recent use cases, such as AI datasets\cite{Malhotra24}, suggest the need for faster proof generation over increasingly large constraint sets. This paper explores efforts in optimizing zero-knowledge proof generation, specifically PlonK, using GPUs, for applications that are not amenable to solution via {\it incrementally verifiable computing} (IVC). 

Our intention with this work is to improve ZK performance by introducing fundamental changes in the low-level architectural structure of proof generation. These changes affect performance curves for most current proof
 generation algorithms. In particular, this paper explores the trade-offs between increased parallelization on GPUs and the resultant memory management problems that occur due to this parallelization. We implement a parallelized PlonK in order to analyze these trade-offs and make arguments concerning the nature of data flows within the proving protocol. Our results lead us to consider multi-GPU environments to maximize performance for large circuit sizes. Such environments introduce issues of communication, efficient hardware management, data partition, and data
 transfers. This paper further discusses the use cases that motivate multi-GPU implementations, their cost models, and the use of {\it remote direct memory access} (RDMA) for data transfer among GPUs. 
\end{abstract}

%%
%% The code below is generated by the tool at http://dl.acm.org/ccs.cfm.
%% Please copy and paste the code instead of the example below.
%%


%%
%% Keywords. The author(s) should pick words that accurately describe
%% the work being presented. Separate the keywords with commas.

\maketitle

\section{Introduction}
\label{sec:intro}
Zero-knowledge succinct arguments of knowledge, or zk-SNARKs, are cryptographic primitives through which a prover $\rho$ may efficiently convince a verifier $v$ of the correctness of some computation $\lambda$. This is achieved by providing a succinct argument of $\lambda$'s validity $\pi$, revealing no further information about $\lambda$~\cite{ZK88}. Such constructions are extremely beneficial in various application domains.  These range from privacy applications, as discussed,\, e.g., in ~\cite{ZcashWhitepaper}, to cases where zk-SNARKs are used
for their succinctness and thus easy correctness checking, as in zk-rollups.  Because of the computational complexity of proof generation, significant attention has been paid to proof-generation efficiency by subdividing tasks.  These include incrementally verifiable computing~\cite{Mina,ZKfolding}, and the aggregation of proofs to distribute easily segmented subtasks~\cite{ZKfolding,rollupReview23}. In this paper, we focus on zk-SNARKs, specifically PlonK\cite{Gabizon2019PlonKPO}, applied to use cases that are sub-dividable into small proofs. We elaborate on these and other application domains in Section~\ref{sec:background}, in Table~\ref{tab:zk_apps}. 

The widespread adoption of zk-SNARKs and other argument-of-knowledge technologies is limited by these performance barriers. Particularly, these challenges lie in the balancing of three key metrics: verification time, proof-generation time, and proof size. Various approaches to constructing arithmetization schemes and implementing underlying cryptographic primitives (i.e. hashing and elliptic curves) provide different balances of these three aforementioned metrics~\cite{Thaler23}. 

Scaling zero-knowledge is a prominent issue within the cryptographic space. Various solutions center either around decreasing algorithmic complexity in ZK systems~\cite{zkTree23,ZKfolding} or efficiently implementing existing systems~\cite{MSMGPU22,GZKP,PipeZK}. The latter approach has seen significant advances in the past several years, due to competitions that
focus on engineering scalable ZK proving systems, such as ZPrize\footnote{\url{https://www.zprize.io/}}. Many solutions focus on common computations shared among proving systems including multi-scalar multiplication (MSM), the number theoretic transform (NTT), and other hard computations. These serve as large bottlenecks in the proof generation process and increasing efficiency in them would have an extreme impact on the power of ZK systems~\cite{PipeMSM}. 

While this paper holistically considers the entirety of the proving process, it focuses on computation time for proof generation as a key metric in our GPU optimizations. Within the proof-generation process, we focus on multi-scalar multiplication (MSM), which is at the core of many zero-knowledge proving systems and is highly suitable for parallelization. The application domains we cite necessitate the creation of ZK proofs over hundreds millions of constraints ($>2^{28}$). Constraint sets of this magnitude lead to extremely high computational complexity and demand for large amounts of memory.

In this paper, we describe out implementation of the PlonK protocol into an NVIDIA A10 and V100 environment using the CUDA platform and Thrust library. Through this process, we observe bottlenecks in memory management stemming from difficulties in optimizing within a heterogeneous CPU-GPU computing architecture~\cite{Roofline}. The memory challenge is shown to be sufficiently serious that a well-configured CPU system can outperform our 
GPU-based implementation despite PlonK being highly parallelizable.

Our work demonstrates the importance of carefully considering not only parallelism but also memory management when choosing an execution strategy for ZK-proof implementations.
We also provide a foundation for further research efforts into parallelizing larger ZK-proof constructions, not only within a GPU, but also across multiple GPUs (and CPUs). Such a system would require a cost model and partitioning algorithm to determine the most efficient partitioning and strategic delivery of data at runtime. Such a ZK-proof implementation would leverage data transfer
between processors using remote direct memory access (RDMA), a method of data transfer that does not interrupt the remote processor across a network of machines. We further expand upon this in Section 7. 

The general outline of the rest of the paper is as follows.
Section 2 provides the necessary background to zk-SNARKs, specifically the PlonK constructions and relevant research that motivates our contributions. Section 3 describes our implementation process and highlights critical development decisions. Section 4 details results from our proving system, and Section 5 analyzes these results and connects them to particularly notable areas of improvement within the MSM. Section 6 explores challenges in optimization of heterogeneous computing, our motivation for pursuing non-IVC based proofs, and brings to light several potential promising improvements in proof generation. Section 7 describes future work motivated by Section 5, concluding our paper.
\section{Background}
\label{sec:background}
Since the introduction of the Groth16 protocol~\cite{Groth:2016}, zk-SNARK protocols have rapidly increased in number. In all of these protocols, certain commonalities arise including the MSM operation on which
we focus in this paper. More generally, three broad stages represent the life-cycle of any SNARK construction:

\begin{itemize}
    \item Setup: the process in which parties engage in a {\it multi-party computation} (MPC) to create a cryptographic reference that secures all future computations. There exist zk-SNARKs with trusted~\cite{Gabizon2019PlonKPO,Groth:2016} and also untrusted setups~\cite{Dark}. PlonK uses a trusted setup.  A key advanage of PlonK is
    its required use of only a single, universal setup, compared to Groth16 which requires a separate setup for each execution being proved.
    \item Proof Generation: the process of generating an argument of knowledge for subsequent verification.
    \item Verification: the process of verifying the argument of knowledge.
\end{itemize}

While this paper focuses on the proof-generation process, we briefly discuss the trusted setup and verification processes as trade-offs made during proof generation. 

During the trusted setup, parties collaborate in an MPC setting to generate a reference string, often referred to as a {\it common reference string} (CRS) or {\it structured reference string} (SRS). This reference string is used to hide universal commitments, which create the structure for the cryptographic guarantees provided later in the proof generation process. Creating a reference string is either circuit-specific (i.e. program specific) or universal. Circuit-specific constructions, like Groth-16, require a new trusted setup for each specific computation to be proved. While this is impractical for typical large-scale ZK applications, Groth16 provides the benefit of a comparatively smaller proof size and shorter prover computation time. Most widely-adopted ZK systems have universal trusted setups, which allow for the creation of a single SRS that can be used for all proofs thereafter. 

The verification stage must be efficient because, in most applications, a proof is verified by multiple parties. For PlonK and Groth16, the proofs are succinct and can be verified in constant time. 
Setup-free proof systems, based on zk-STARKs\cite{STARKS18}, have the obvious benefit of no setup phase, but suffer from asymptotically worse verification time (linear or logarithmic).
Looking ahead, STARKs, being hash-based, offer quantum safety.

\subsection{PlonK}
PlonK~\cite{Gabizon2019PlonKPO} introduced the construction of a prover with a universal SRS via the arithmetization scheme Plonkish. PlonK and its variants have since become a standard in zk-SNARK-based constructions; and it reasonably approximates the state of zero-knowledge SNARK technology. 

Central to PlonK and important to the efficiency of the PlonK system is its arithmetization scheme: Plonkish. For our purposes, Plonkish can be described as a protocol acting upon polynomial constraints over a finite field. These polynomial constraints restrict our circuit to specific input/output requirements. Each gate represents one instruction in a computation, and its associated polynomial must evaluate to zero for the gate condition to be satisfied. We can represent these polynomials with a sparse matrix that eventually transforms into a polynomial of degree $3n + 5$ for a circuit with $n$ gates.

To get a sense of how large even relatively simple circuits
can be, we note that a circuit that demonstrates that a prover knows three terms in a Pythagorean triple contains about 12 constraints, resulting in an MSM with at least 41 summations of elliptic curve points\cite{PlonkByHand}. More circuit sizes are detailed in Table~\ref{tab:zk_apps}. These summations are difficult to compute because, while each coefficient in this polynomial is a scalar, each term associated with a coefficient is a point within a finite field. This field is instantiated over an elliptic curve, with points often above 200 bits in size. The MSM can be described as follows:
\begin{equation}
    MSM = \sum_{i = 0}^{n-1}G_i*x_{i}
\end{equation}
where each elliptic curve point, an element of our field, is represented by $G_i$ and our scalars are represented by $x_{i}$. The scalars $x_i$ are too large to be processed as hardware-level integers, so they must be split into smaller \textit{limbs}, denoted as $x_{i,j}$.  The number of bits, $c$, in a limb is called the \textit{window size}.  The result is that for a $b$-bit scalar, we have $K=b/c$ windows, and within each window
there are $2^c$ possible scalar values. Thus, we write the MSM as a sum over 
each of the $K$ windows:
\begin{equation}
    MSM = \sum_{i = 0}^{n-1} \sum_{j = 0}^{K-1} G_i*x_{i,j}*2^{cj}
\end{equation}
where the $2^{cj}$ term comes from the need to adjust each partial result based on the position of each limb in the full scalar.
\begin{table}[htb]
        \tiny
        \caption{\textbf{Examples of ZK Applications and Relative Constraint-Set Sizes}}
        \label{tab:zk_apps}
        %scale table to single column
        \resizebox{\columnwidth}{!}{%
        \begin{tabular}{p{0.1\columnwidth} p{0.305\columnwidth} p{0.07\columnwidth}}
            \toprule
            \textbf{Example} & \textbf{Description} & \textbf{Size} \\
            \cmidrule{1-3}
            \textit{ZK-Rollups}\tablefootnote{ https://research.protocol.ai/sites/snarks/} & ZK-Rollups are a substantive application of ZK technology, and are virtually limitless in size. & $>2^{27}$ \\
            \cmidrule{1-3}
            \textit{Private Transactions}\tablefootnote{https://research.protocol.ai/sites/snarks/} & ZK powered private transaction systems provide a solution to transacting privately on a public ledger.  & $>2^{26}$ \\
            \cmidrule{1-3}
            \textit{Trustworthy AI}\tablefootnote{https://www.certik.com/resources/blog/advanced-formal-verification-of-ZK-proof-verifying-a-zk-instruction}  & AI accountability in the realms of data provenance and distribution can be approached readily from a ZK perspective. & $>2^{23}$ \\
            \cmidrule{1-3}
            \textit{Supply Chain}  \tablefootnote{https://www.gate.com/learn/articles/succinct-perfecting-the-infrastructure-for-zk-proof-supply-chains/2325} & ZK solutions offer fair-trade guarantees about consumer goods via traceability and authenticity. & $>2^{23}$ \\
            \cmidrule{1-3}
            \textit{Anonymous Voting}\footnote{https:/link.springer.com/chapter/10.1007/978-3-031-72244-8\_7} & Anonymous voting systems are realizable through ZK systems which protect user identities. & $>2^{20}$ \\
            \cmidrule{1-3}
            \textit{Health AI Model Verification}\cite{VerfAIMedPooyan} & Proof that a specific health diagnosis used a specific approved AI diagnostic model & $>2^{30}$\\
            \bottomrule
        \end{tabular}%
            }
            \end{table}
            
\subsection{MSM Acceleration Techniques}
The existence of applications with large circuit sizes and the difficulty of performing operations in elliptic-curve
groups over large finite fields has led to a variety of 
algorithms and techniques to speed up computation.
In this section, we detail algorithms and techniques that are widely adopted among virtually all proving systems due to their efficiency in facilitating the MSM.

\begin{figure}
    \centering
    \includegraphics[width=0.95\linewidth]{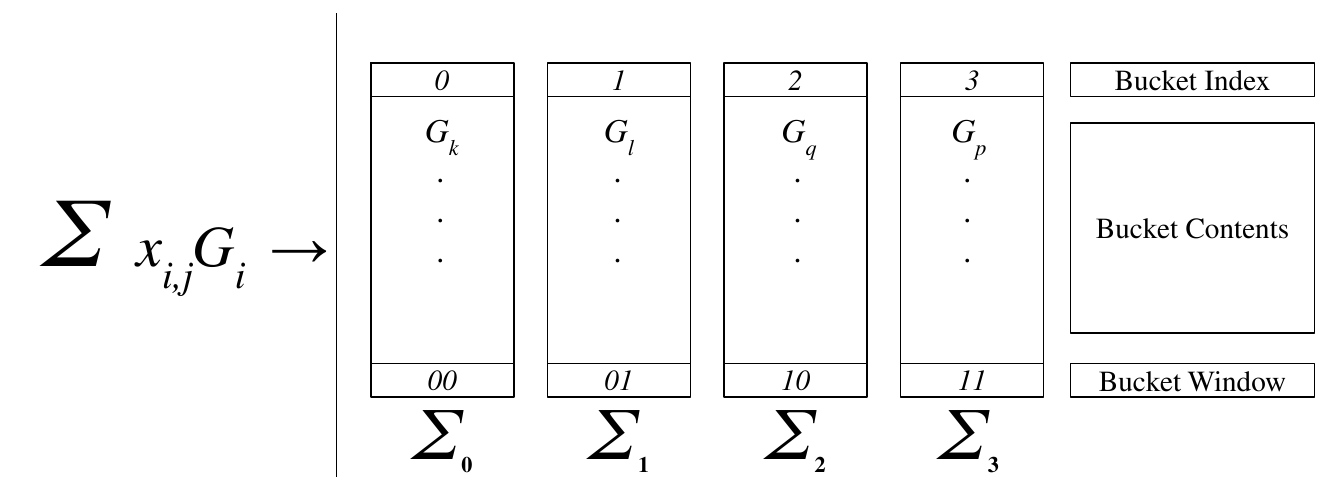}
    \caption{Bucket Accumulation in Pippenger's Bucket Method}
    \label{fig:bucket_acc}
\end{figure}

First, and most notably, is Pippenger's bucket algorithm~\cite{Bernstein2002PippengersEA}. The primary focus of this algorithm is to eliminate the large scalar multiplications associated with each point and replace them with many smaller point additions and multiplications. The algorithm directly addresses this change through the introduction of windowing. During the bucket accumulation step, the scalars of the computation are split bitwise into windows of arbitrary size chosen by the developer. Each window-sized scalar is then  associated with a specific bucket based on its value, as shown in Figure ~\ref{fig:bucket_acc}. Points are then assigned to buckets based on the scalar by which the point is to be multiplied. After the values are assigned, the sum of the points in each bucket is computed, and each of those sums is multiplied by total by the scalar associated with that bucket. After that, for each window, the bucket sums are themselves summed. The window-sums are then scaled via a bit shift to correspond to the size of the original scalars. Finally, the totals of the scaled window-sums are accumulated to obtain the final result.

Although the main goal of this algorithm is to break up large scalars into smaller values to avoid costly large point multiplication, the implementation comes with an extremely parallelizable approach. This parallelism is shown through the steps of breaking up the scalars into buckets within windows and through the accumulation of those buckets and windows. Each of these processes can be done at once without having the complication of collisions between threads on a CPU or GPU. The difficulty comes with how to adjust these operations accordingly for the most efficient method to balance compute times with data transfer. More recently, compelling work has suggested assigning GPU cores to buckets, with threads still being assigned to windows \cite{distmsm}.

\begin{figure}
    \centering
    \includegraphics[width=0.9\linewidth]{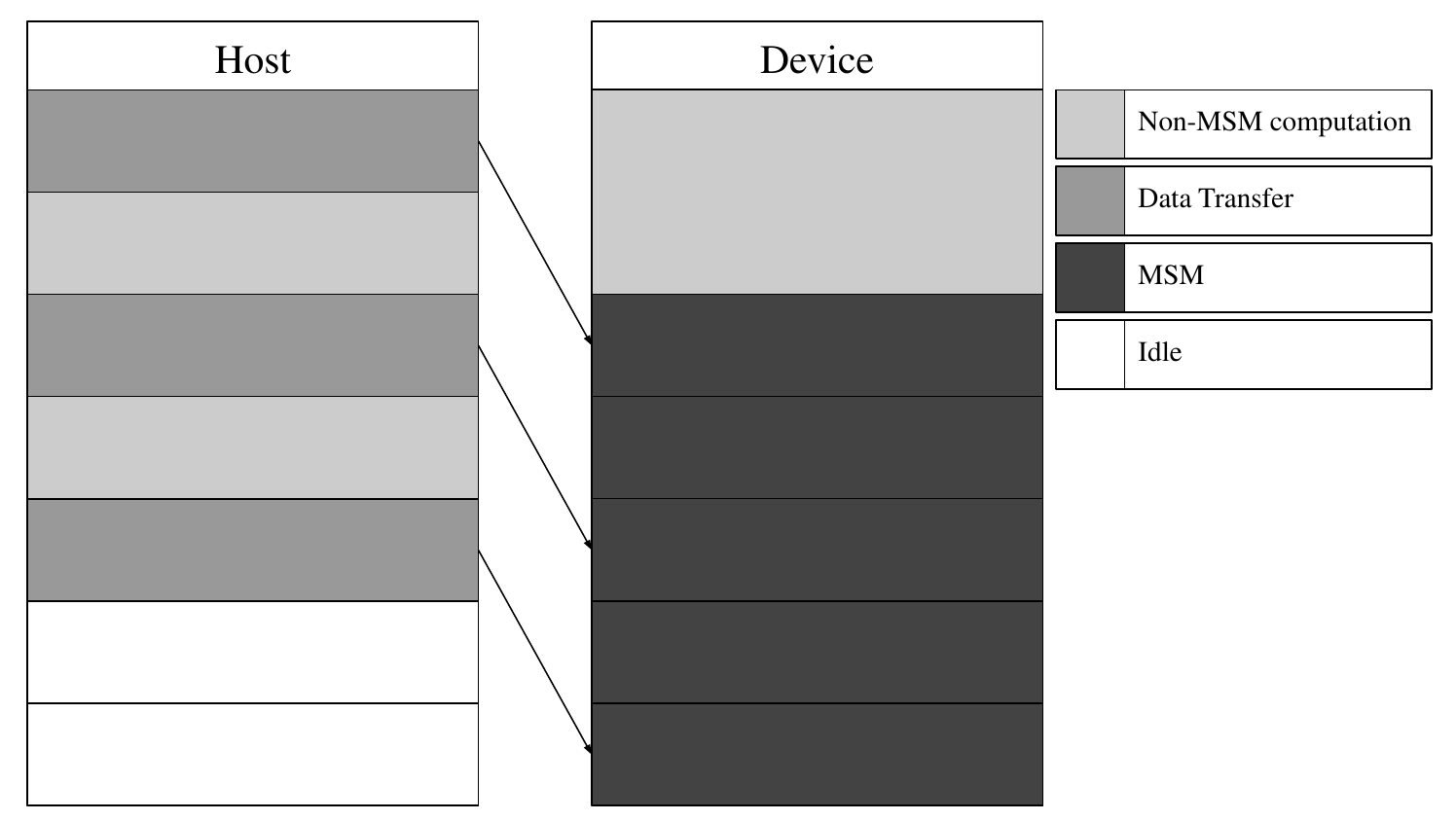}
    \caption{Diagrammatic representation of memory interleaving data transfers in between respective host and device transfers/computations.}
    \label{fig:memintl}
\end{figure}

Stemming from Pippenger's bucket method, and another common avenue for accelerating MSM dominated SNARK proofs, is memory interleaving~\cite{MSMGPU22}. Memory interleaving takes advantage of the multistreaming technique to simultaneously transfer data loads from host to device while performing computation on device, as depicted in Figure ~\ref{fig:memintl}. This is a very common approach to achieving runtime speedups within the MSM, and is critical to understanding this paper's implementation.

\section{Methodology}
\label{sec:methodology}
The parallelization work was performed within the NVIDIA ecosystem on A10 and V100 GPUs using the CUDA platform and Thrust library. Our work further builds upon Aztec's Barretenberg library, which implements the PlonK protocol. 
Detailed in previous works~\cite{ZKVDBS23,TalTurboUltraPlonKFeb23}, the issue of this parallelization work concerning the PlonK proving system is highly non-trivial. The proving system's various computational bottlenecks and the existence of several sequential computations within those methods \cite{Pipelonk26,sumcheck2025} give rise to  problems in achieving full parallelization. In particular, critical problems exist in the realm of host-device data transfer. In this section, we detail our approach to developing an initial parallel implementation of the PlonK proving system aimed at providing a platform for measuring memory requirements and motivating our decision to use Thrust. We describe the steps needed for a full
implementation in Section~\ref{sec:improvement}.

\subsection{CUDA Kernels and Modularization}
Our experiments focused directly on MSM computation with the limbs input directly as a large array. This array contained a series of big-ints represented as four 64-bit limbs, that is, each element of the array was allotted four indices to fit the size of the 254-bit scalars and field elements used in the MSM. The array also contained pre-allocated and unoccupied indices, reserved to allow for an easier-to-use indexing scheme. The indexing scheme is used in the bucket value reduction computations. The experiment's data were loaded in this manner to avoid the need for multiple host-to-device data transfers before and throughout the on-device computations. 

For the parallel execution of our implementation, we used CUDA, a general purpose API library and extension of C++ made by NVIDIA for working with GPUs. CUDA abstracts the underlying GPU via the PTX virtual-machine architecture.  Within the CUDA extension of C++, kernels can be made that are functions compiled to run specifically on the GPU to utilize the thousands of threads a GPU can support. Several of these kernels are used in our implementation of the MSM. For these kernels, we chose a sequential implementation of individually highly parallel CUDA launches relying upon the CUDA-Thrust framework. This is done to perform the multitude of sum-reductions, a parallel computing computation critical to the overall performance of Pippenger's algorithm~\cite{Bernstein2002PippengersEA} for computing the MSM.

For our work, we chose the BN-254 curve\cite{BN-254}, though other curves are also used in practice.
Our BN-254 curve is employed using coefficients modulo a prime close to the size of $2^{254}$, requiring us to perform arithmetic operations on integers of at least 254 bits in length (``field elements''). We used 4 extended-precision 64-bit ``add'' instructions in the PTX assembly, plus a small cost to check if we overflowed the modulus (+4 instructions to correct any overflow). Given a statically known prime, an efficient solution is to use a technique known as Montgomery multiplication\cite{EdMSM2022}. In our case, this allows computing multiplication modulo of the aforementioned prime using only 64-bit multiplications and bit shifts. Having no direct step of modulo a 254-bit prime eliminates the need to track higher-order bits or implement a bignum modulus. Our implementation uses 32 multiply-adds, 16 additions, and one check for overflow correction. This is almost entirely implemented directly using PTX assembly’s support for extended-precision integer arithmetic with carry propagation. 

These changes allow us to use the Thrust library in our bucket accumulation stage of the MSM by delegating the prefix sum code to the Thrust abstractions, improving both performance and simplicity. Thrust’s implementation, backed in turn by CUDA Unbound (CUB) as part of the CUDA C++ core libraries, uses a highly optimized algorithm incorporating decoupled lookback to  calculate efficiently the prefix sum over a very large number of inputs\cite{nvidia-merrill}. The result is a highly optimized and adaptable implementation that scales efficiently to higher input sizes, never needing to spawn a kernel manually. The dimensions of the execution grid are no longer tied to the bucket size, as the new algorithm simply processes all points and then scatters the output to the correct locations as necessary.

\subsection{Hardware and Testing}
In order to analyze systems with circuit counts ranging from $2^{15}$ to $2^{25}$, careful selection of GPU hardware was important. We focused on finding units that we expected would lend themselves well to proof generation. We remind the reader that these constraint-set sizes are entirely reasonable in terms of applications and are less than the scale of the largest applications  that could benefit from ZK technology, as discussed in Section~\ref{sec:background}. 

Workloads consisted of a series of benchmarks using this GPU implementation as well as previous related works\cite{talthesis} using the Volta Tesla V100 GPU and the Ampere A10 GPU. The main differences between these chips are how they handle memory transfers. The Tesla V100 is a chip has been developed for high-performance computing applications while the A10 is a chip has been developed for efficiency in AI applications. Note that the A10 GPU has roughly 1.8x more CUDA cores available than the V100. Another difference is how these GPU implementations handle memory and bandwidth. Specifically, the Volta architecture we used has 16 gigabytes of memory\footnote{The V100 offers a 32GB VRAM configuration, but that was not available for this work.}  with 900 gigabytes per second of bandwidth~\cite{NVIDIAvolta}. The Ampere A10 architecture supplies 24 gigabytes of memory with 600 gigabytes per second of bandwidth~\cite{NVIDIAa10Datasheet}. These specifications show a 1.5x improvement in memory and a 0.66x decrease in bandwidth for the Ampere A10 architecture in comparison to the Volta architecture. 

Both chips were chosen to be benchmarked as they both provide different computing capabilities that may prove optimal for the sub-problems of the ZK proof computation. Specifically, an increased number of CUDA cores in the A10 allows for more streaming multiprocessors(SM). The V100 allows for better allocation per streaming multiprocessor, memory bandwidth, and more capable SM scheduling due to Volta’s redesigned memory hierarchy. We chose these two architectures because they both have unique advantages that lend themselves well to proof generation. Furthermore, they handle memory transfers differently~\cite{NVIDIAvolta,A100GPU}, and studying how they react to high memory loads is of acute interest to future work in this space, discussed in Section~\ref{sec:future}. 

\section{Results}
\label{sec:results}

\begin{figure}[t]
    \centering
    \includegraphics[width=0.95\linewidth]{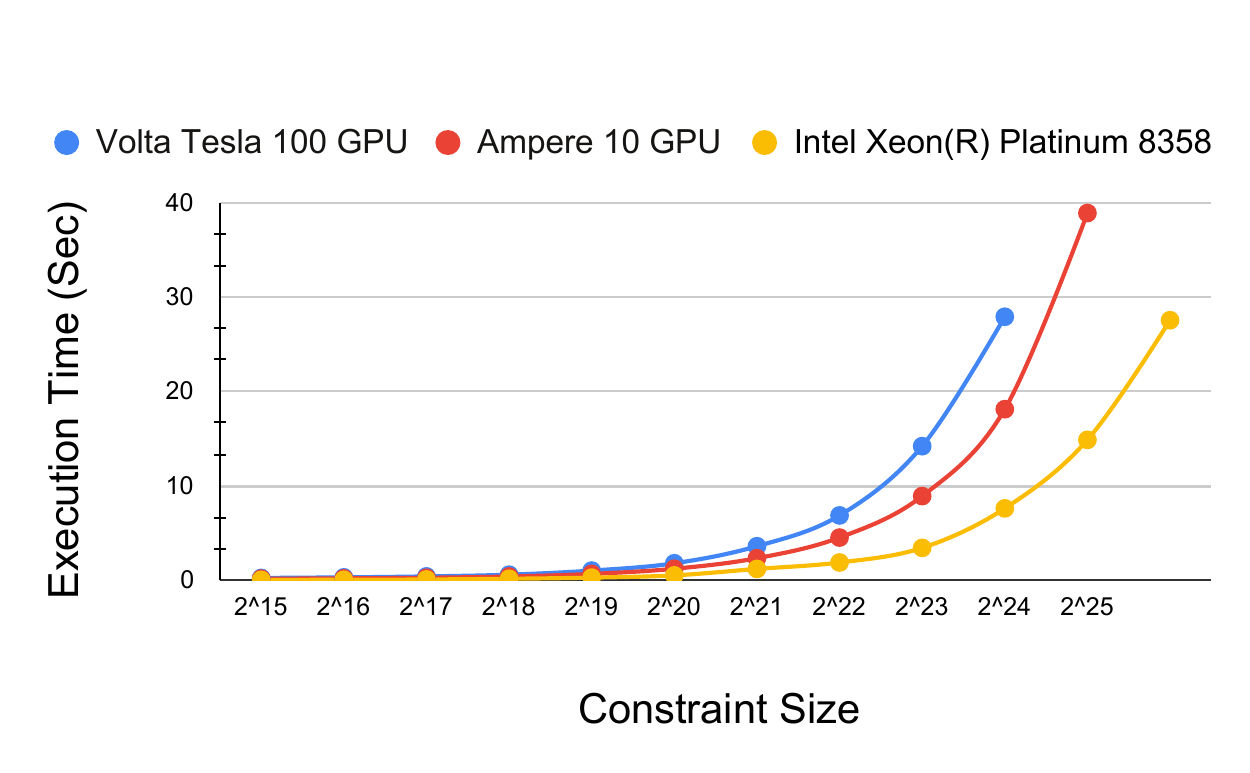}
    \caption{Execution Size Vs. Constraint-Set Size and Processor}
    \label{fig:exec_time}
\end{figure}

\begin{figure}[t]
    \centering
    \includegraphics[width=0.95\linewidth]{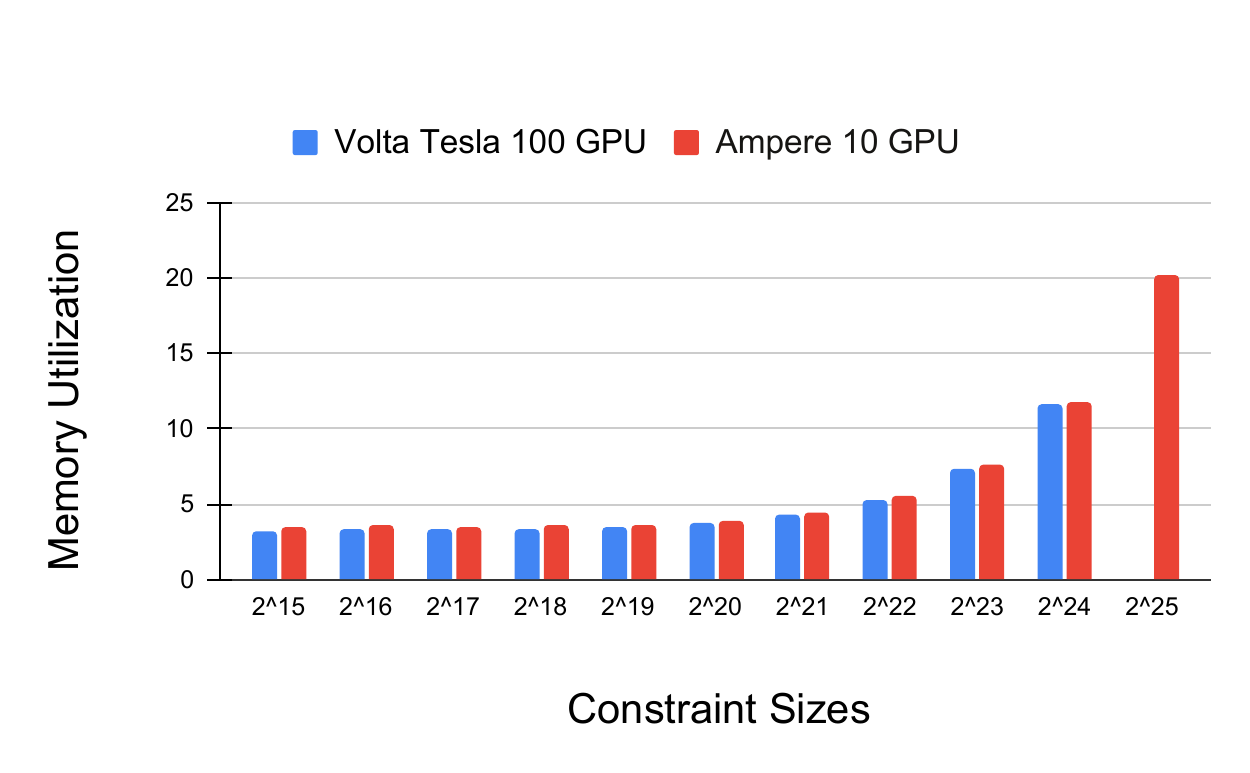}
    \caption{Memory Utilization Vs. Constraint-Set Size}
    \label{fig:mem_util}
\end{figure}

We present our examination of the parallelized proving system in the following section. Special care was placed on the study of GPU memory utilization and the role of data transfers between host (CPU) and device (GPU) as a bottleneck, especially in how the computation's GPU parallelism utilization is bound by memory rather than by compute capability. Ultimately, our results show that the performance bottleneck lies in memory congestion and data transfer. Figure~\ref{fig:exec_time} shows that execution time as a function of the size of the constraint set grows at a dramatic rate once GPU memory limits are hit. Figure~\ref{fig:mem_util} shows that, for each GPU system, the performance barrier is hit when memory utilization approaches 100\%. In contrast, running on a CPU-only platform shows significantly better performance. That is, our ability to exploit the full degree of GPU parallelism was limited by memory issues. These metrics formulate the basis of our approach to the problem of expanding this work to multiple GPUs, further detailed in Section~\ref{sec:future}. 

Note that validation of the MSM result is of utmost importance in verifying that the PlonK system ran correctly and is a non-trivial problem. In our implementation, correctness was achieved by using result references of operations done previously on the elliptic curve and associated points that we used. These references were received directly from Aztec, the developer of the native PlonK proving system within their Barretenberg library.

%\subsection{Ampere and Volta Architectures}
%dropped subsection since there is only one!

We analyze the performance of the PlonK system on an instantiated bare-metal NVIDIA V100 and A10 Tensor Core units. We consider the performance based upon (1) scaling with respect to constraint-set sizes and (2) overall memory utilization.

Focusing on constraint-set size is natural in this analysis, as the computation size directly influences the number of constraints in the system~\cite{maxthesis}. Generally, within the implementation, we see Ampere units receiving a slightly larger memory utilization throughout proof generation compared to Volta units. We observe that Ampere 10 finishes execution in slightly less than 20 seconds up to $2^{24}$ constraints, where the Volta Tesla 100 finishes its computation in slightly less than 30 seconds up to the same constraint-set size. 

The Ampere and Volta architectures saw nearly identical and near-constant memory utilization up until $2^{21}$ constraints, where exponential growth begins. This growth is insurmountable on these platforms, and past $2^{24}$ constraints, the Volta architecture failed to finish the computation. While the Ampere GPU was able to outperform the Volta architecture, it too failed at $2^{25}$ constraints. We observe similar failures to the Ampere results for $2^{25}$ constraints on the Volta hardware, detecting VRAM utilization of 20 GB.

We note similar, yet not equal, memory requirements in the two architectures despite disparities in the compute time for large constraint-set sizes. 

These results are consistent with the improved hardware capabilities that the Ampere architecture has compared to Volta, namely in memory. Specifically, Ampere architecture A10 has a 1.5x improvement in memory compared to the Volta architecture, indicated in the hardware specifications,  resulting in a huge difference in the hardware demand with increasing numbers of constraints. This heightened load on the hardware past the point of $2^{21}$ constraints is observed in compute time measurements. Notably, this corresponds to a markedly low performance compared to the CPU implementation, where memory was not a limiting factor. 

In Section~\ref{sec:analysis}, we consider why the GPU implementations failed to outperform the CPU implementation regardless  of the number of constraints.

\section{Analysis}
\label{sec:analysis}
The results in previous section show a clear problem with our
initial focus on parallelism, since
GPU memory usage and data transfer between host and device memory are the bottlenecks in the computation, overwhelming the benefits
of parallelism. GPU memory usage becomes a significant limiting factor with a large number of constraints due to the mapped bucket indexing used in the implementation, and the need to allocate room for each possibly filled bucket before the first kernel launch. This implementation choice optimizes inter-kernel performance using existing Thrust reduction methods to achieve the greatest possible parallelism on a per-kernel basis. A more GPU-architecture amicable implementation could exploit host-device memory transfer interleaved with computation and shared memory transfer (as opposed to global memory transfer). 
Memory interleaving is only beneficial as long as we have VRAM space. Consequently, we propose that despite improvements from shared memory implementations, larger constraint-set sizes will necessitate the exploration of optimizations beyond interleaving. 

Data transfer from host to device and vice versa proved to be a large execution time bottleneck. We opted for a chosen bucket indexing scheme which requires an initial transfer to GPU for sum reduction. While this results in decreased efficiency, more complicated indexing schemes typically result in larger memory overhead.~\cite{MSMGPU22}. 

The option of implementing a less complicated indexing scheme is
a reasonable alternative that was beyond the scope of this work. It is worth noting that in future implementations with regard to large amounts of MSM-input constraints, multiple data streams from host hardware to device hardware could potentially reduce the bottleneck~\cite{MSMGPU22}. Nevertheless, as described in section~\ref{sec:improvement}, these may require RDMA operations. Mentioned earlier in section~\ref{sec:methodology}, the use of Thrust both capitalizes on an easily optimized bucket running sum technique but also delays the generation process due to the necessity of converting between host types (i.e. primitive arrays) and Thrust types such as Thrust vectors. 

Also worthy of considerations are the problems that a Thrust-reliant implementation introduces. Thrust is particularly well suited for optimizing running sums at runtime and is a natural choice for highly intensive running-sum loads. One particular issue of our Thrust implementation is the lack of ability for the Thrust interpreter to handle large memory transfers between device (the GPU) and host (the CPU), which are required for every Thrust call. It is important to note that this memory congestion is not a Thrust-specific issue. While memory bottlenecks may be slightly exacerbated with a Thrust implementation, the scale of the MSM is still the primary hurdle for any host-device transfer process that may not be completely amenable to multi-streaming techniques. 

The conversion of host arrays and primitive storage types necessitates their adaptation into thrust-readable vector conversions.
In the MSM, we see primitive array of sizes $4 * 2^{26}$ or greater~\cite{Barretenberg}. 
The existence of types as large as 712 bits requires the deconstruction of elliptic curve points into multiple segments. As a result, despite optimal parallelism within kernels, parallel PlonK fundamentally suffers from overly congested memory traffic, especially within the MSM computation.
In Section~\ref{sec:results}, we see complications from massive conversions of the order of $> 2^{26}$ size arrays that are frequently converted from host-to-device and device-to-host. Because of these reasons, we see that Thrust is likely not an optimal choice for implementing running sums with large workloads, such as the one found in the computation of the MSM. These particular data structures, necessitated by the MSM and used throughout the various kernels and host processes, greatly affected the final computation.
We discuss this further in
Section~\ref{sec:future} and in \cite{nusratthesis}.

% It should be noted, however, that this congestion is surmountable. We see evidence that through memory interleaving, we can alleviate memory transfer congestion while performing MSM computations to incorporate larger amounts of streaming multiprocessors \cite{MSMGPU22}. Moreover, we bring attention to the issue of memory and the management of data and I/O streams in CPU-GPU memory transfer. Because our implementation does not utilize multiple I/O streams in MSM computation, we see that we are bound by the global memory of the GPU rather than by the DRAM allocated to the CPU. 

\section{Alternative Techniques}
\label{sec:improvement}
Before moving forward, it is important to note that our utilization of Thrust in the implementation of Aztec's Barretenberg (PlonK) does not affect the algorithmic structure of PlonK. It also does not utilize multiple I/O streams to interleave MSM computations on the GPU with memory transfer as seen in other implementations, such as cuZK\cite{MSMGPU22}. Our parallel implementation paired with this work primarily aims to bring attention to the true bottleneck in tackling the problem of parallelizing the multi-scalar multiplication of points in large finite fields. In looking forward to improving upon our implementation, attention must be paid to sufficiently leveraging the parallelism on the GPU through increasing the pressure placed on I/O streams. This will then cause a greater number of streaming multiprocessors than what is currently used.

With this in mind, we discuss several optimizations that may dramatically improve the scaling potential of zk-SNARKs. Primarily, these areas concern the promising sectors of ZK research surrounding folding and incremental verifiable computation (IVC) in general. This scheme presents a recursive approach to proof generation of computations. It is also worth noting that at times IVC protocols avoid SNARKs, and even arguments of knowledge in general. These eliminate the need for a trusted setup or an FFT, and recent works show smallest known prover times and verifier overheads~\cite{ZKfolding,Mangrove}.

While this incremental approach is promising, we see three unique reasons to believe non-recursive SNARKs still signify an important aspect of verifiable computation: 
\begin{enumerate}
    \item It is difficult to parallelize IVCs due to their linear nature. Consequently, computations that are generally approached from a parallel computing lens suffer when verified incrementally. There exist tree-based implementations of IVC that allow certain degrees of parallelism through the use of tree-based recursive protocols, but fully exploiting the nature of certain embarrassingly parallel computations will always be inherently limited in IVC-based schemes. 
    \item The folding schemes of Mangrove~\cite{Mangrove}, and Sangria~\cite{Sangria} both rely upon Plonkish arithmetization schemes to perform commitments to the verifier. We want to highlight Sangria, which utilizes a zkSNARK - PlonK - to provide stronger arguments for succinctness and zero-knowledgeness in their incremental computation. 
`    \item The problem of the MSM still presents itself in nearly every IVC scheme in the literature~\cite{ZKfolding,Mangrove,Sangria,ZcashWhitepaper,spartan,ZKfolding}, especially those which utilize folding. As described in Section~\ref{sec:intro}, the MSM dominates prover time in SNARKs like PlonK. As a result, improvements in SNARKs via reductions in MSM compute time and efficient memory sharing also stand to benefit in the space of folding schemes.
\end{enumerate}

Beyond the above the reasons listed above, the dominant consideration is that 
folding schemes and recursive proof constructions are, in general, not applicable to problems that have a large volume of input data. More specifically, there exist problems that require verification not only of a statement of the latest computation over the latest subset of data but also verification of a statement over the latest subset of data along with previous subsets of data. Large proofs that fall into this category are of particular interest in the application domain of privacy-preserving healthcare software applications, such as the analysis of healthcare images, such as MRIs \cite{VerfAIMedPooyan} with varying file sizes for different health-related images ranging from 131 KB to 30 MB. Other examples of large ZK-proof applications that do not meet the input requirements of folding schemes or recursive construction include:
\begin{enumerate}
    \item Financial Data Analysis with an interest in financial structuring for the purposes of evasion.
    \item Verification of an outsourced computation's integrity, especially in cases where the computation is built upon a black-box model.
\end{enumerate}
These examples highlight the problem set for which every additional proof may require data that have already been verified via IVC. 

If parallelization of folding schemes can be applied to IVC as a whole, and especially in the context of optimizing zkSNARKs found in certain folding schemes, then very promising offers of scalability arise. These scalability directions are even more beneficial in the context of parallelizable IVC's, such as Mangrove~\cite{Mangrove}. While IVC is a promising domain, aforementioned downfalls indicate it is not a silver bullet, and we still must explore other realms for ZK acceleration.
%(https://www.researchgate.net/figure/File-sizes-of-images-from-different-imaging-modalities_tbl1_26831435#:~:text=CT%2C%20MR%2C%20and%20the%20US,%5B19%5D%20.%20...)%

\section{Future Work}
\label{sec:future}
As applications with increasingly large numbers of  constraints emerge, it becomes imperative that proof-generation techniques continue to advance to meet demand. In this section, we introduce several areas of future work, motivated by the results and observations we have presented in this paper and by emerging proving technologies that may be consequential for zk-SNARK generation schemes.

First, the absolute need for GPU memory suggests a multi-GPU approach, even if all resulting parallelism cannot be exploited. The goal is to run the MSM computation at GPU speed to optimize proof-generation latency. While a multi-GPU system seems straightforward, it introduces various other issues. Specifically, data-transfer bottlenecks would be even more prevalent than what we faced in our implementation for this paper. While we needed to move data between CPU and GPU, any partitioning of computations among GPU systems requires even more movement of memory. Recent developments, such as Dist-MSM\cite{distmsm}, have further laid the foundation for multi-GPU implementations. Dist-MSM tackles these issues by limiting how often points and scalars are transferred, performing heavy computations locally, and aggregating partial results in a master node.

Although our implementation reached maximum constraints of $2^{25}$, there are applications with a greater number of constraints, approaching a billion ($2^{30}$). Furthermore, potential use cases extend beyond those referenced, and the need for large constraint-set sizes is poised to grow. This presents the need for a multi-GPU approach and, more specifically, a cost model for data-transfer that accounts for all transfers among computers in a parallel computing cluster. Beyond data transfers, the nonuniformity of the MSM must also be considered as partitioning data requires attention. Seeing as the ultimate goal is scalability and latency, it is critical that, for each computation, the data arrive in a timely manner to optimize the full computing platform. Additionally, it is advantageous to allocate total memory usage in an efficient manner among multiple GPUs.  For the MSM, this suggests that we must seek parallelization that is consistent with efficient data movement. 

Note that folding is absent from this list of future work. For the aforementioned reasons, we chose to overlook IVC schemes for the time being in favor of bolstering memory issues through the above techniques in the short term. However, we still support the idea that IVC schemes will be a leading solution in accelerating zero-knowledge for proof tasks that are segmentable into subtasks. 

That being said, within the scope of multiple GPUs, we clearly see an issue arise in memory sharing, which already exists between host and device on a single GPU. Thus, we again emphasize the value of developing a cost model for the size of the input constraint-set size to the hardware requirement. It is consequential that this model consider the cost of memory transfers, any trade-offs, and the ideal partitioning of data among a multi-GPU system. This will be dampened by the potential use of RDMA within memory transfers, a promising new route of development for memory intensive programs~\cite{GPUDirect}. 

We see a compelling and emerging direction for scaling zero-knowledge proofs. Specifically, the eventual implementation of a multi-GPU model\cite{nusratthesis} in conjunction with the use of RDMA to handle particular proofs of correctness that cannot be incrementally verifiable.

\begin{acks}
This work is supported by Oracle Cloud and related resources provided by the
Oracle for Research program, and by gifts from Steel Perlot and  Google.
A Lehigh CORE grant also provided support.  
We thank dePaul Miller, our colleague in the SSS Lab at Lehigh, for providing his
GPU expertise.

Finally, we acknowledge our colleagues in the Scalable Systems and Software Research group (\url{sss.cse.lehigh.edu}) and in Lehigh Blockchain (\url{blockchain.cse.lehigh.edu}).
\end{acks}

%because max is 8 pages excluding references
%%
%% The next two lines define the bibliography style to be used, and
%% the bibliography file.
\bibliographystyle{ACM-Reference-Format}
\bibliography{blockchain}

\end{document}